\documentclass[aps,prl,nofootinbib,reprint]{revtex4-1}

\usepackage{graphicx}
\usepackage[utf8]{inputenc}
\usepackage{amssymb}
\usepackage[colorlinks,pdfusetitle]{hyperref}
\hypersetup{allcolors=[rgb]{1,0.56,0}}

\newcommand{\lsim}{\lesssim}
\newcommand{\gsim}{\gtrsim}

\newcommand{\eq}[1]{Eq.~(\ref{#1})}

\newcommand{\ord}[1]{\mathcal{O}{(#1)}}
\newcommand{\beq}{\begin{equation}}
\newcommand{\eeq}{\end{equation}}
\newcommand{\bea}{\begin{eqnarray}}
\newcommand{\eea}{\end{eqnarray}}
\newcommand{\eps}{\epsilon}

\newcommand{\orcid}[1]{\href{https://orcid.org/#1}{#1}}
\newcommand{\e}[1]{\times10^{#1}}

\begin{document}

\title{An Attractive Scenario for Light Dark Matter Direct Detection}

\author{Hooman Davoudiasl}
\email{hooman@bnl.gov}
\thanks{\orcid{0000-0003-3484-911X}}
\author{Peter B.~Denton}
\email{pdenton@bnl.gov}
\thanks{\orcid{0000-0002-5209-872X}}
\author{Julia Gehrlein}
\email{jgehrlein@bnl.gov}
\thanks{\orcid{0000-0002-1235-0505}}
\affiliation{High Energy Theory Group, Physics Department, Brookhaven National Laboratory, Upton, NY 11973, USA}

\date{\today}

\begin{abstract}
Direct detection of light dark matter (DM), below the GeV scale, through electron recoil can be efficient if DM has a velocity well above the virial value of $v\sim 10^{-3}$.  We point out that if there is a long range attractive force sourced by bulk ordinary matter, {\it i.e.}~baryons or electrons, DM can be accelerated towards the Earth and reach velocities $v\sim 0.1$ near the Earth's surface.  In this ``attractive scenario," {\it all DM} will be boosted to high velocities by the time it reaches direct detection apparatuses in laboratories. Furthermore, the attractive force leads to an enhanced DM number density at the Earth facilitating DM detection even more. We elucidate the implications of this scenario for electron recoil direct detection experiments and find parameters that could lead to potential signals, while being consistent with stellar cooling and other bounds.  Our scenario can potentially explain the recent excess in electron recoil signals reported by the XENON1T experiment in the $\sim$~keV energy regime as well as the hint for non-standard  stellar cooling.       
\end{abstract}

\maketitle

\section{Introduction}

A variety of astronomical and cosmological observations have established that the Universe contains a substance of little, if any, interaction with ordinary matter made of atoms.  This substance, dark matter (DM), comprises about $25\%$ of the cosmic energy budget, which translates to about $85\%$ of all matter in the Universe \cite{Aghanim:2018eyx}.  Not much is known about non-gravitational properties of DM, due to its elusive nature.  Given the diversity of particles and forces that constitute the ``visible" sector encoded in the Standard Model (SM) of particle physics, it is reasonable to consider whether DM resides within a ``dark sector" that comprises a number of new states and forces that only feebly interact with the SM.

There has been a significant experimental effort over the last few decades to detect DM in the laboratory.
This effort has been matched by intense theoretical research directed at DM phenomenology.  The questions surrounding the physics underlying electroweak symmetry breaking in the SM and its extensions led to an early focus to look for DM around the weak scale $\sim 100$~GeV.  The lack of evidence for new physics near that scale, from high energy and precision experiments, together with null signals for weak scale DM in a variety of searches, has provided motivation to expand the experimental and theoretical efforts to lower masses, where new challenges arise.  In the realm of direct detection, going to lower DM masses means smaller available energies in collisions of DM particles with detector target material, which requires lower detection energy thresholds and controlling backgrounds that could overwhelm the signal.

Searches designed for weak scale DM are mostly focused on looking for nuclear recoil signals.  However, direct detection of light DM, below the GeV scale, motivates looking for electron recoil signals.  To see this, let us consider some rough estimates.  The typical virial velocity of DM near the solar system is $v\sim 10^{-3}$.  For heavy DM masses, $m_{DM}\sim 100$~GeV, this corresponds to a nucleus of mass $m_N\sim 10$~GeV recoiling with momentum 
of order $q \sim m_N \,v \sim 10^{-2}$~GeV and energy $E_R \sim q^2/(2 m_N)\sim 10$~keV.  Now, 
if we consider sub-GeV DM masses, 
say $m_{DM}\sim 0.1$~GeV, we see that the momentum transfer is $q\sim m_{DM}\, v \sim 100$~keV and the nucleus would recoil with energy $E_R\sim$~eV, which is well below the $\gsim$~keV threshold of such experiments.  

The above situation can be improved if one looks for electron recoil signals. To see this, note that electrons in atoms are delocalized over length scales of order Bohr radius $a_0 \sim (\alpha\, m_e)^{-1}$, where $\alpha\approx 1/137$ is the fine structure constant and $m_e \approx 511$~keV is the electron mass.  Thus, the typical momentum of the electrons in the atom is $q_0\sim 1/a_0$ and the electron velocity is hence $v_e \sim \alpha$, which is much larger than the virial velocity of DM.  Nonetheless, the recoil energy of the electron will be $E_R\sim q_0^2/(2 m_e)\sim 10$~eV.  Hence, detection of a signal in electron recoil in an experiment with $\gsim$~keV energy threshold requires velocities near $v\sim 0.1$, which is well above the escape velocity from the Milky Way, $v_{esc}\sim \text{few}\times 10^{-3}$, severely suppressing the expected abundance of any such DM particles in the halo population.

Given the above situation, to look for typical DM in the sub-GeV regime, one needs to devise experimental techniques with detection thresholds $\ll$~keV \cite{Crisler:2018gci,Abramoff:2019dfb,Barreto:2011zu,Arnaud:2020svb,Agnese:2018col} (for novel ideas see \cite{Schutz:2016tid,Hochberg:2015pha,Hochberg:2017wce,Bloch:2016sjj}).  
Alternatively, one could investigate new DM models that could be detected in the current class of large scale experiments like XENON1T or the planned next generation searches such as XENONnT and LZ \cite{Aprile:2015uzo,Aprile:2017aty,Mount:2017qzi}, with thresholds near the keV scale.  In fact, there have been some ideas put forth in recent years where a fraction of the DM can have velocities $\gg 10^{-3}$, due to originating from decays of a more massive DM state \cite{Huang:2013xfa,Agashe:2014yua,DEramo:2010keq} or else due to interactions with energetic particles, such as cosmic rays \cite{Bringmann:2018cvk,Ema:2018bih}.  In these schemes,
typically only a small fraction of DM could be boosted to higher velocities.  

In what follows, we will propose a novel scenario, where {\it all DM} reaches the Earth at velocities 
$v\sim 0.1$ without leading to severe depletion of the Galactic halo or requiring interactions that are in conflict with
laboratory or astrophysical
bounds.  The basic idea is that there could be a long range {\it attractive} force that acts on DM and ordinary matter, but with unequal strengths \cite{Davoudiasl:2017pwe}.  
This force need not have a range -  
set by the inverse mass of the boson that mediates it - that far exceeds the size of the Earth.  The matter in the Earth, baryons or electrons, can then source a long distance potential that accelerates DM particles towards the Earth once they get close to it.   We will show that for appropriate choices of parameters, one could attain velocities $v\sim 0.1$ for all DM particles that reach the surface of the Earth, which to a very good approximation characterizes the location of typical DM experiments.  We will also introduce a short range interaction that mediates DM-electron scattering; as an example we will choose a light vector boson that kinetically mixes with the photon, {\it i.e.}~the dark photon \cite{Holdom:1985ag}. 

As will be discussed in the following, our scenario opens up a new possibility for detection of light DM at existing and planned experiments that use electron recoil with thresholds of $\sim$~keV.  Interestingly, XENON1T has observed an excess in electron recoil events that is significant at $\sim3.3$ $\sigma$ \cite{Aprile:2020tmw}.  After presenting  the central idea of our work, we will  discuss the possibility of explaining this potential signal of DM in our scenario, while maintaining agreement with stringent bounds from stellar cooling considerations. 

For a recent work that takes advantage of a long range interaction to avoid stellar bounds in explaining the XENON1T excess, but in a different model and context, see Ref.~\cite{DeRocco:2020xdt}.  The utility of a high velocity DM population for explaining the excess was emphasized in Ref.~\cite{Kannike:2020agf} early on; see also Ref.~\cite{Alhazmi:2020fju}.
For some recent works that have also 
considered a light vector boson as a mediator for the reported XENON1T excess, see for example Refs.~\cite{Alonso-Alvarez:2020cdv,Boehm:2020ltd,Baryakhtar:2020rwy,An:2020bxd,Lindner:2020kko,Baek:2020owl,Okada:2020evk,Choi:2020kch,AristizabalSierra:2020edu,McKeen:2020vpf,Fornal:2020npv,Bally:2020yid}.

Next, we will introduce the long range force described above.  We will then introduce an example of a short range interaction that will be necessary for detectable scatterings of DM on electrons.  

\section{Long Range Interactions}
\label{sec:long range}

We propose a long range interaction between ordinary matter (baryons or electrons) and DM, as was suggested in Ref.~\cite{Davoudiasl:2017pwe}.
Depending on its type, such an interaction can lead to an attractive force between DM and atoms.  For concreteness, let us assume that the force carrier is an ultralight 
boson $\phi$ of mass $m_\phi \sim 3\times 10^{-16}$ eV  which is compatible with superradiance limits of ultralight bosons \cite{Baryakhtar:2017ngi}.  This gives $\phi$ a range $\sim 100 \,R_\oplus$, where $R_\oplus \approx 6.4\times 10^3$~km is the radius of the Earth.  Assuming that, for example, nucleons and DM particles $\chi$ couple to $\phi$ with strengths $g_n$ and $g_\chi$, respectively, the entire Earth sources a potential 
for $\phi$ of the form
\beq
V(R)\approx - g_n g_\chi\frac{N_\oplus}{4\pi R}\,,
\label{V}
\eeq
at a distance $R$ from the center of the Earth, where $N_\oplus \sim 10^{51}$ is the number of nucleons in the Earth.  Here, we have assumed that $R \lsim m_\phi^{-1}$, so that the interaction is not Yukawa suppressed.  If $g_n g_\chi > 0$, then the above potential leads to an attractive force.  

Searches for new long range forces 
lead to a very stringent constraint $g_n\lsim 10^{-24}$ \cite{Schlamminger:2007ht,Fayet:2017pdp}.  However, on length scales of order $100 R_\oplus$, there are no severe 
constraints on interactions of $\phi$ with DM and one could have \cite{Davoudiasl:2017pwe}  
\beq
g_\chi \lsim 4\times10^{-6} \left(\frac{m_\chi}{1~\text{MeV}}\right)^{3/4}\,.
\label{gchi}
\eeq
We then have
\beq
V(R_\oplus) \sim -0.01\text{ MeV}\, \left(\frac{g_n}{10^{-26}}\right) \left(\frac{g_\chi}{10^{-7}}\right)\,.
\label{VRe}
\eeq
The above provides $E_{KE}=-V(R_\oplus)$ of kinetic energy for every particle coming from infinity, with the usual $v\sim 10^{-3}$ virial velocity, after falling down the potential well approaching the surface of the Earth.  The velocity at $R=R_\oplus$ is then given by 
\beq
v(R_\oplus) \sim \sqrt{\frac{-2V(R_\oplus)}{m_\chi}}\,.
\label{vRe}
\eeq
So, for $m_\chi=1$ MeV and reference values in \eq{VRe}, we find $v\sim0.14$.

The long range force also leads to an enhancement of the number density of the DM particles at the Earth 
similar to the particle density enhancement around a black hole \cite{Peirani:2008bu}.
The density is increased by 
the ratio of the DM velocity at the Earth over the DM velocity in the solar system  
\begin{equation}
 r_v\approx v_{\text{final}}/v_{\text{initial}}\,.
 \label{eq:rv}
\end{equation} 
In our model, we have $r_v\sim 100$. This enhancement can be understood as an increased cross section of the Earth which requires  the range of the long range force to exceed the impact parameter of this interaction.

\section{Short Range Interactions}
\label{sec:short range}
The potential that accelerates DM, introduced in the previous section, does not mediate electron scattering processes that can be observed 
in DM direct detection  experiments.  Hence, we need to introduce another interaction, of much shorter range, to have detectable signals.    
As an example for such an interaction we will focus on the case of a light dark photon mediator $A_D$, with mass $m_D$, which mixes kinetically with the photon, described by the Lagrangian 
\begin{equation}
    \mathcal{L}\supset \frac{\epsilon}{2}F^{\mu\nu}F_{D \mu \nu} -\frac{m_D^2}{2} A_{D\mu}A_D^{\mu} + i\, e_D A_{D\mu} \overline{\chi}\gamma^\mu \chi\,,
\end{equation}
where $F_{(D) \mu\nu}$ is the field strength tensor for the (dark) photon and $e_D=\sqrt{4\pi \alpha_D}$ is the dark photon coupling.

In our scenario, due to the high velocity of DM reaching the detector, we can use the ``free electron" approximation.  In this case, for the energies and momentum transfers of interest we can ignore the atomic binding energies, as long as we only consider the outer shell electrons. For the case of xenon atoms, used as target material in the current and planned large scale DM detectors, this 
corresponds to electrons in the $n = 4$ and 5 levels, for a total effective charge of $Z_{eff} = 26$.  In an approximation where the electrons are treated as free and 
initially at rest, we find the differential cross section for DM electron scattering (for some relevant formalism, see for example Ref.~\cite{Essig:2015cda}) 
\beq
\frac{\text{d} (\sigma_e \,v)}{\text{d} E_R} =\frac{8 \pi \,m_e\, \alpha\, \alpha_D \epsilon^2 \, Z_{eff}}{v\, (2 m_e E_R + m_D^2)^2}\Theta(2\mu_{\chi e}^2v^2/m_e-E_R)\,,
\label{diff-sigmav}
\eeq
where the electron recoil energy is given by $E_R = |\vec{q}|^2/(2 m_e)$, with the magnitude of the three-momentum transfer denoted by $|\vec{q}|$.
The step function provides the kinematic limit. 

We can get the total cross section by integrating \eq{diff-sigmav}. In order to regulate the infrared behavior of the cross section, we will introduce a threshold energy 
$E_{th}$, below which events are not registered by the experiment.  We then find,
\beq
(\sigma_e \, v) = \frac{16 \pi\, \alpha\, \alpha_D \epsilon^2 Z_{eff}  (\mu_{\chi e}^2 v^2 - m_e\, E_{th}/2)}{v\, (2 m_e \,E_{th} + m_D^2)(4 \mu_{\chi e}^2 v^2 + m_D^2)}\,,
\label{sigmav}
\eeq
where $\mu_{e \chi}$ is the reduced mass of the electron-DM system, $1/\mu_{\chi e} \equiv 1/m_e + 1/m_\chi$.  In the above, the maximum recoil energy is given by $E^{max}_R = 2 (\mu_{\chi e}^2/m_e) v^2$.  

Using \eq{sigmav}, we can write down the expected rate per detector mass and year,
\beq
\frac{\text{d} R}{\text{d} t \,\text{d} M} =  n_T\,  n_\chi (\sigma_e v)\,,
\label{dRdtDM}
\eeq
where $n_T = 6.02 \times 10^{23}\, \text{g}^{-1}/A$ is the number of target atoms per gram, with $A$ the target atomic mass, and $n_\chi=r_v \rho_\chi/m_\chi$ is the number density of DM particles; the DM energy density is $\rho_\chi \approx 0.3$~GeV cm$^{-3}$ \cite{Tanabashi:2018oca} and the enhancement of the number density $r_v$ from \eq{eq:rv}.

In the above, due to the nearly uniform boost of all DM to $v\sim 0.1$ at the detector, we may approximate the DM velocity distribution by a delta function
\begin{equation}
f(v) \approx \delta [v - v(R_\oplus)]\,, 
\end{equation}
near the surface of the Earth.  

\begin{figure}
    \centering
    \includegraphics[width=1\linewidth]{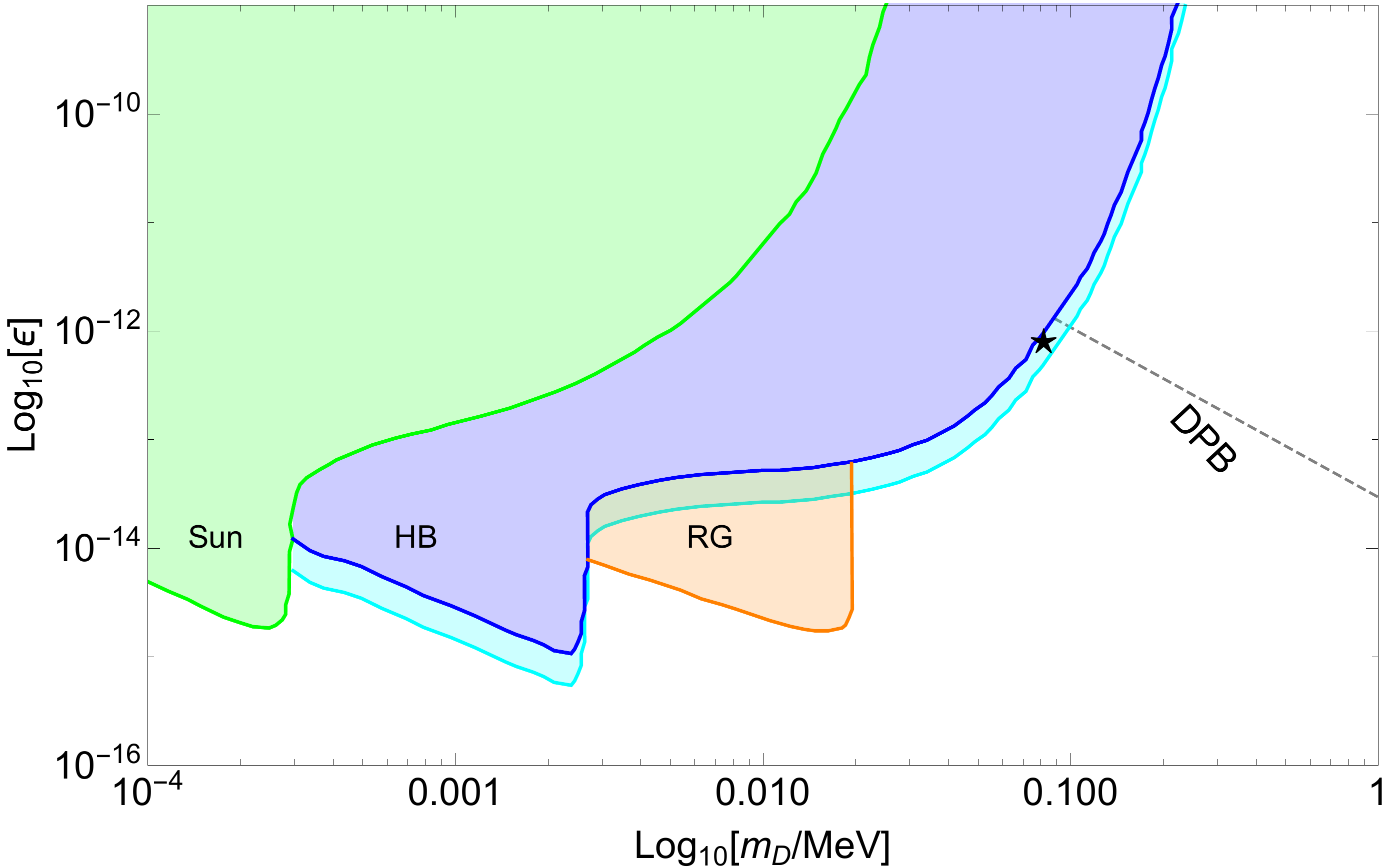}
    \caption{ Constraints on dark photon mass vs.~kinetic mixing.  The constraints are adapted from Ref.~\cite{Essig:2013lka}. The light blue region represents the HB cooling hint from Ref.~\cite{Giannotti:2015kwo}.    The black star represents the value of the benchmark point from tab.~\ref{tab:fiducial} which can explain the XENON1T excess.  }
    \label{fig:results}
\end{figure}

 For light dark photons with $m_D\lesssim 10$ keV the cross section is independent of the dark photon mass whereas for large dark photon masses the signal rates depends on $m_D^{-4}$.
These results  need to be compared to constraints on the mass of a dark photon and its kinetic mixing taken from \cite{Essig:2013lka}. We will restrict ourselves to the region between $100~\text{eV}<m_D<1 $ MeV where the decay of the dark photon into SM fermions is not kinematically allowed.
In the region between $m_D>1$ eV up to 0.1 MeV strong constraints on the kinetic mixing come from stellar cooling of the Sun, of stars in the horizontal branch (HB), and for red giants (RG).
We note that there is a slight hint of new physics in HB cooling measurements  which could be explained by a dark photon for parameters shown in fig.~\ref{fig:results} \cite{Giannotti:2015kwo}.

Between $m_D\sim $ 0.1 and 1 MeV constraints on the kinetic mixing  from the diffuse photon background  (DPB) apply. However, in our model these constraints can be evaded by assuming a light dark fermion that would allow prompt {\it invisible} decays of the dark photon.  This may seem to lead to conflict with the number of relativistic degrees of freedom allowed during Big Bang Nucleosynthesis (BBN).  However, for values of $\epsilon \lsim 10^{-11}$ of interest in our work, the dark sector and the SM sector would not be in equilibrium and the dark sector could be much ``cooler" than the visible sector, making it unconstrained by these considerations.  To see this, note that the rate for $e^+ e^- \to \gamma A_D$, as an example, is roughly given by $\alpha \epsilon^2 T$, which at the BBN temperatures of $T\sim \ord{\rm MeV}$, is $\ll H(T)$, where the Hubble scale is set by $H(T) \sim T^2/M_P$, with $M_P\approx 1.2 \times 10^{19}$~GeV the Planck mass. At higher temperatures the decoupling of the two sectors is enhanced and for $T\lsim 1$~MeV electrons have annihilated away, suppressing thermalization processes.

Around $m_D\approx 0.1$ MeV where stellar cooling measurements lose sensitivity 
we find a sweet spot which allows for kinetic mixings which can simultaneously explain the HB hint and XENON1T. 
This benchmark point is also allowed by the general constraints on dark photons as well as by constraints from Supernovae and BBN on light DM interacting via a dark photon \cite{DeRocco:2019jti,Chigusa:2020bgq}.

To summarize, the fiducial parameters of the model are shown in table \ref{tab:fiducial}.
\begin{table}
\centering
\caption{The fiducial parameters of the model.
The first three parameters are related to the long range interaction while the last four are related to the short range $\chi-e$ scattering interaction.}
\begin{tabular}{c|c|c}
$m_\phi$ & $g_n$ & $g_\chi$ \\\hline
$3\times 10^{-16}$ eV & $10^{-26}$ & $10^{-7}$ 
\end{tabular}\\\vspace{0.1in}
\begin{tabular}{c|c|c|c}
$\alpha_D$ & $m_D$ & $\eps$ & $m_\chi$ \\\hline
$2\times10^{-6}$ & 0.08 MeV & $8\times 10^{-13}$ & 1 MeV
\end{tabular}
\label{tab:fiducial}
\end{table}

\section{Discussion}
\label{sec:discussion}
We note that XENON1T has recently reported a slight excess of electron recoils in the few keV range \cite{Aprile:2020tmw}.
While backgrounds such as tritium could possibly explain the excess, these explanations appear to be disfavored, yet more investigation may be necessary for a firm conclusion.

\begin{figure}
\centering
\includegraphics[width=\columnwidth]{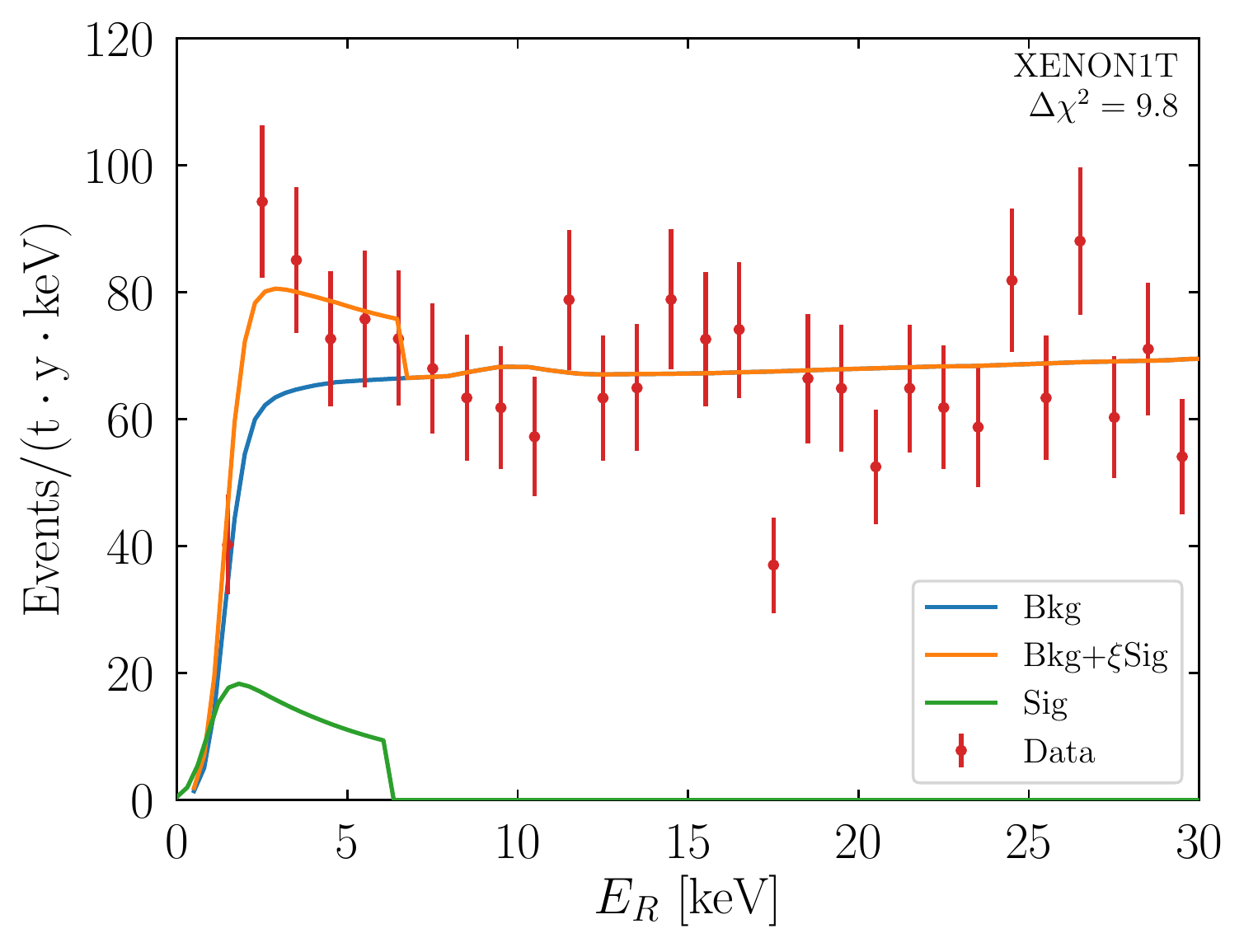}
\caption{The XENON1T data in red and their best fit background model in blue.
Green is our best fit signal curve and the orange curve is the background plus the signal times the XENON1T efficiency.
The best fit point for $m_\chi=1$ MeV is at $\alpha_D\eps^2=1.5\e{-30}$, $m_D=0.082$ MeV, and $v=0.12$ at which point we find $\Delta\chi^2=9.8$ compared to the background only.}
\label{fig:data}
\end{figure}

We performed a fit of the parameters to the data as shown in fig.~\ref{fig:data}.
To do so, we computed the differential cross section from \eq{diff-sigmav}, multiplied it by $n_Tn_\chi$, defined  following \eq{dRdtDM} including the $r_v$ factor, and applied the efficiency $\xi$ given in Ref.~\cite{Aprile:2020tmw}.
For a test statistic we computed a simple $\chi^2$ function considering only the error bars in the data points.
We then marginalized this function over the DM velocity $v$, the dark photon mass $m_D$, and the normalization parameters $\alpha_D\eps^2$ assuming $m_\chi=1$  MeV and that we are in the non-relativistic limit while maintaining the full dark photon propagator.
We find that our model is preferred over the background only hypothesis with $\Delta\chi^2=9.8$.

We can see that a key feature of the model is not only a suppression of events at low energy due to the dark photon mass, but also at high energy due to the sharp velocity distribution.
Unlike many other explanations of the XENON1T data, we anticipate that the spectrum would fall off fairly sharply at higher recoil energies.
Our best fit parameters are $\alpha_D\eps^2=1.5\e{-30}$, $m_D=0.082$ MeV, and $v=0.12$ which has a test statistic $\chi^2=36.6$ compared with the background only hypothesis which is $\chi^2=46.4$.

\begin{figure}
\centering
\includegraphics[width=\columnwidth]{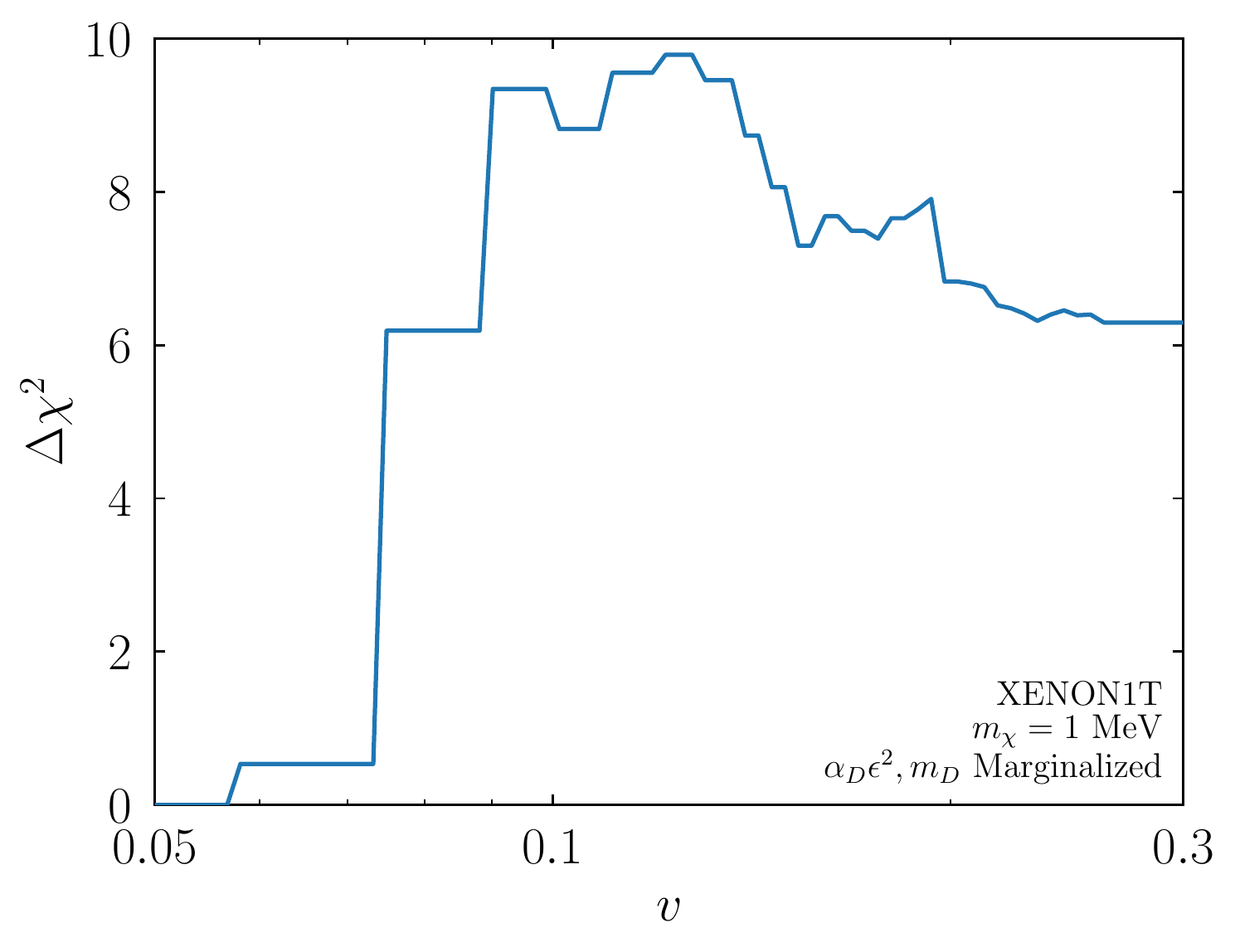}
\caption{The preferred region of velocity after marginalizing over the normalization style parameters such as  $\alpha_D$, and $\eps$ as well as $m_D$.
The sharp nature of the plot is due to the binning of the data.}
\label{fig:velocity}
\end{figure}

To further understand the dependence of the parameters on the data, we show two interesting $\chi^2$ projections for the XENON1T data.
In fig.~\ref{fig:velocity} we show the velocity projection where $m_\chi=1$ MeV and the other parameters are marginalized over.
We see that $v\sim0.1$ is preferred.
For smaller velocities the DM does not have enough kinetic energy to have an effect above XENON1T's threshold.
At high velocities an improved fit is found, but it slightly overestimates the signal at larger recoil energies.
In practice the velocity distribution is not truly a delta function as we have modeled it here and some of DM would have higher velocities which would make the suppression at higher recoil energies a bit softer.
Nonetheless we anticipate that this is a small effect.

\begin{figure}
\centering
\includegraphics[width=\columnwidth]{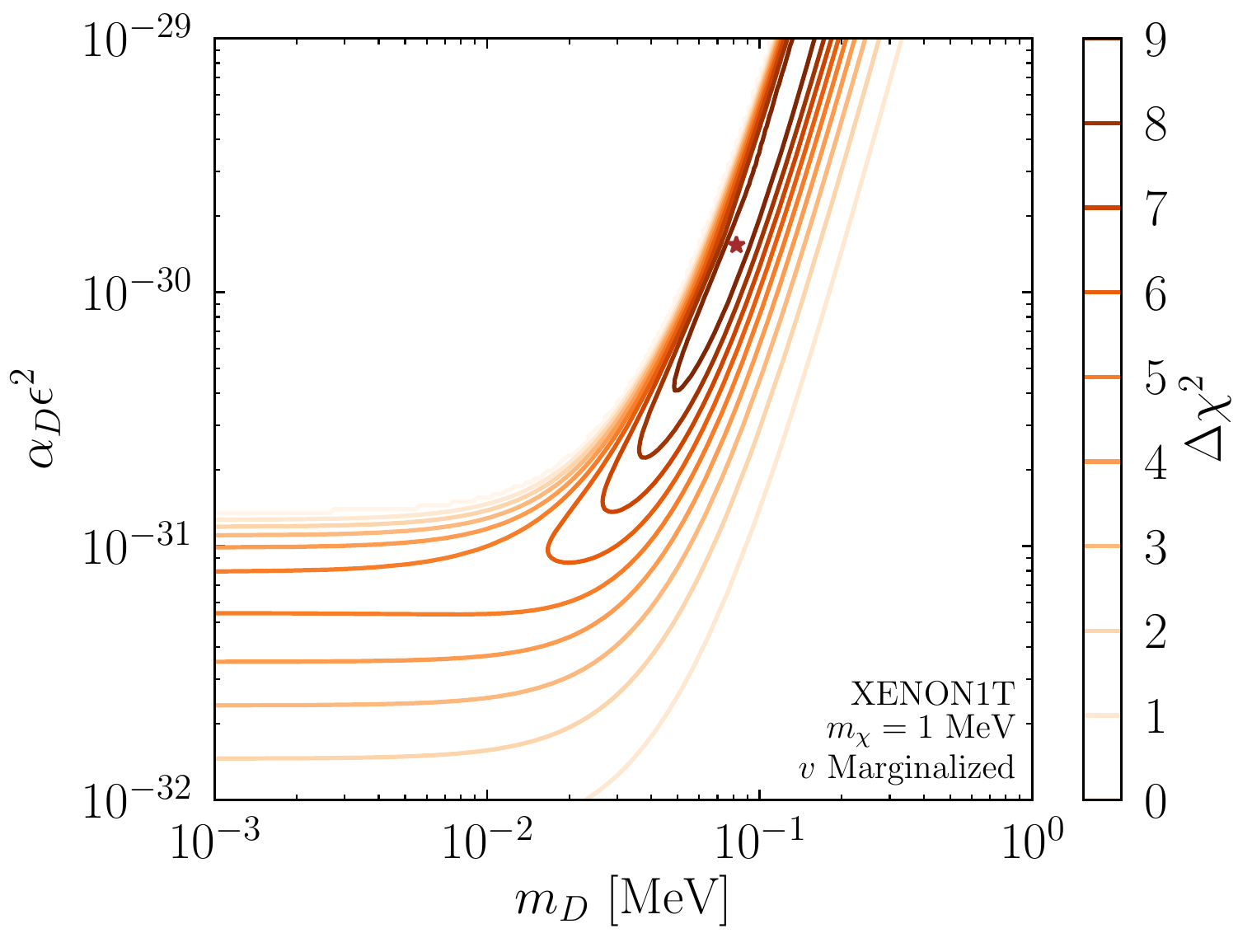}
\caption{The parameters that are preferred by the XENON1T data; the center is preferred by the data, the bottom right returns to the SM, and the top left produces too big of a signature and is strongly ruled out.
We compute the $\chi^2$ test statistic between the data and the background plus the signal rate times the efficiency function.
This is compared to the $\chi^2$ between the data and the background alone.}
\label{fig:region}
\end{figure}

Next we investigate the parameters of the short range interaction in fig.~\ref{fig:region}.
We see that the best fit point is for $m_D\sim0.1$ MeV and $\alpha_D\eps^2\sim10^{-30}$  which is for example satisfied for  $\alpha_D=2\times 10^{-6}$, $\eps=8\times 10^{-13}$, and $m_\chi=1$ MeV which is also consistent with other bounds shown in fig.~\ref{fig:results}. We also note that for the preferred values of $\alpha_D$, $m_\chi$ and $m_D$ the DM self-interaction cross section $\sigma\approx  4 \pi \alpha_D^2m_\chi^2/m_D^4$ satisfies the approximate constraint $\sigma/m_{\chi} \lesssim 1~\text{cm}^2/\text{g}$ \cite{Tulin:2013teo}. 
The innermost region of fig.~\ref{fig:region} is maximally preferred at $\Delta\chi^2>9$, the region to the top left is significantly disfavored as it over-predicts the signal, while the region to the bottom right is generally consistent with the background only hypothesis as it predicts no additional events.
The best fit region continues up to larger normalizations, although these become ruled out from other constraints as one must  dial up the couplings ($\alpha_D$ or $\eps$).

Beyond the current large-exposure low-threshold experiments, this model can be, in principle, tested in other ways.  The attractive nature of the potential in our scenario would provide a nearly radial flux, both up-going and down-going, for DM close to the Earth's surface, a sort of ``dark matter rain," which would lead to significant anisotropy of the signal.  This hypothesis could be tested in experiments that have directional sensitivity \cite{Sekiya:2003wf,Sciolla:2008vp,Daw:2011wq,Miuchi:2012rma,Santos:2013hpa,Battat:2013gma,Cappella:2013rua,DAmbrosio:2014arr,Couturier:2016isu,Hochberg:2016ntt,Kadribasic:2017obi,Rajendran:2017ynw,Budnik:2017sbu,Griffin:2018bjn,Coskuner:2019odd} as the anisotropy is different from both solar neutrinos and the conventional isotropic DM blizzard.

The long range interaction component of this model provides another unique, although difficult to test, prediction.
Confirming a direct detection signal of DM would require multiple independent detections of the signal.
Due to the velocity gain as DM falls into the Earth, this model predicts that the detection rate will be altitude dependent.
That is, we expect a very slightly higher rate at detectors in underground mines such as LZ at SURF which is 1.5 km below the surface than those at the surface such as XENON1T at Gran Sasso.

\section{Conclusions}
\label{sec:conclusions}
We have presented a unique model of dark matter (DM) wherein the Earth provides an attractive force on it due to an ultralight mediator.
While this does not significantly modify the evolution of DM in the Galaxy, this potential does provide a large effect on the velocity distribution near the Earth, in particular by considerably adding to the velocity of 100\% of the DM and yielding a nearly radial flux.
Thus, instead of $v\sim10^{-3}$, all of the DM could have much higher velocities which considerably changes the phenomenology of low target mass recoil experiments such as electron recoils.
In addition, the resultant velocity distribution is highly peaked.
We have included a dark photon sector in our model to provide a testable interaction between DM and electrons.
This model is consistent with known astrophysical, cosmological, and laboratory experiments and possibly explains a tension in stellar cooling data.

Our scenario is testable at low-threshold large-volume DM direct detection experiments such as XENON1T.
In light of the fact that XENON1T has recently seen a tantalizing excess of events at low recoils, we investigated the compatibility of this model with those  data.
We found a good fit to the data for model parameters that are consistent with other bounds.
In addition, this model makes several distinguishing predictions.
Although some would be extremely difficult to test without some rather extreme experiments -- such as a XENON1T like experiment in space or on the moon -- others are much more down to Earth.  In particular, the scenario entails a nearly radial flux of high velocity DM at Earth surface, giving rise to ``dark matter rain," which could be tested in experiments with directional sensitivity.  Also, with future XENON1T data, one can test if the excess has a shape compatible with our prediction shown in fig.~\ref{fig:data}, in particular a suppression at both lower and higher recoil energies, a feature that is not common in many other models.

\begin{acknowledgments}
We thank Garv Chauhan for pointing out a typo in
\eq{vRe}, in an earlier version of this manuscript.
We acknowledge the United States Department of Energy under Grant Contract No.~DE-SC0012704.
\end{acknowledgments}

\bibliography{main}

\end{document}